\documentclass[lettersize,journal]{IEEEtran}
\usepackage{amsmath,amssymb,amsfonts}
\usepackage{algorithm}
\usepackage{algpseudocode}
\usepackage{array}
\usepackage{booktabs}
\usepackage{graphicx}
\usepackage{textcomp}
\usepackage{stfloats}
\usepackage{xcolor}
\usepackage{tikz}
\usepackage{titlesec}
\usetikzlibrary{arrows.meta,positioning,calc}
\usepackage{cite}
\usepackage{fancyhdr}
\fancypagestyle{preprint}{\fancyhf{}\fancyhead[C]{\footnotesize\textsc{Preprint}}}
\usepackage{hyperref}

\graphicspath{{figures/}}
\begin{document}

\title{Towards a Fully Differentiable Ultrasound System for High Frame Rate Plane-Wave Imaging}

\author{Somashekhar~Dehury and Mahesh~Raveendranatha~Panicker
\thanks{Corresponding author: Somashekhar Dehury.}
\thanks{This is a preprint. It has not been peer-reviewed and published in a journal. The proposed source code and a demo has been made publicly available.}
\thanks{Somashekhar Dehury and Mahesh Raveendranatha Panicker are with the Infocomm Technology Cluster, Singapore Institute of Technology, Singapore (e-mail: 2570041@sit.singaporetech.edu.sg; mahesh.panicker@singaporetech.edu.sg).}
\thanks{This work was supported by the Ministry of Education, Singapore, under the Academic Research Fund Tier 1 (GMS 1052), SIT-SINGA Research Scholarship and Tricog Health.}}

\maketitle
\thispagestyle{preprint}

\begin{abstract}
Ultrasound image quality depends on the probe geometry and the transmit and receive processing, however the engineers usually design them separately. The popular ultrasound simulators (Field~II, k-Wave, MUST/SIMUS) compute the radio-frequency channel data without any gradients and hence the design studies must use grid or black-box searches for identifying optimal designs which becomes computationally demanding with the increase in the number of parameters. In this work, we present a fully differentiable end-to-end pipeline for high frame rate plane-wave ultrasound imaging, with synthetic speckle phantoms, an analytical frequency-domain RF simulator, delay-and-sum beamforming and an extensible image-quality objective. The element pitch, the transmit and receive apodization and the global speed of sound are differentiable parameters of one computation graph, and hence one reverse-mode pass is only required to get their exact joint gradients. Studies of individual parameters show that the optimized pitch decreases the grating lobes, and the speed-of-sound autofocus gives clear contrast and resolution gains. Over 15 test phantoms, the optimized pitch increases the cyst gCNR from 0.55 to 0.66 (about 20\%), and the speed-of-sound correction adds a further 0.05. On 49 real CUBDL plane-wave acquisitions from five institutions, the framework finds the global speed of sound with a mean absolute difference of 16.6~m/s from the published corrections and phantom speed of sound.
\end{abstract}


\section{Introduction}
\IEEEPARstart{C}{oherent} plane-wave compounding (CPWC) allows ultrafast ultrasound imaging by trading the number of steered transmits $K$ for frame rate against image quality: the frame rate scales as $1/K$ while contrast and resolution improve with $K$~\cite{montaldo2009}. Traditionally, software parameters, such as steering angles and receive apodization and hardware parameters, such as probe element pitch, have not been jointly optimized as a coupled design problem to improve the image quality. Design choices have traditionally relied on heuristics as the available evaluation tools were non differentiable forward only simulators (Field~II~\cite{jensen}, k-Wave~\cite{treeby2010kwave}, MUST/SIMUS~\cite{garcia2022}). For such non-differentiable forward models, searching the design space meant sweeping grids or running derivative-free optimization, which scales poorly as parameters are added and couples badly across the transmit/receive boundary. If, however, the whole pipeline from RF simulation through beamforming to the image-quality metric, is expressed as a single differentiable program, one reverse-mode pass yields the gradient of image quality with respect to every design parameter, regardless of their number, and the acquisition/probe co-design becomes ordinary gradient-based optimization. This is the idea behind end-to-end differentiable design in computational imaging, where the optics or the sampling pattern is optimized together with the reconstruction~\cite{sitzmann2018,wang2022dO,bahadir2020loupe,suarez2025distilling}. In ultrasound, model-based design has so far targeted the transmit code~\cite{dogan2025coded} or a coding mask~\cite{vandermeulen2020coding}, not the array geometry. In this work, we implement an end-to-end differentiable framework for ultrasound imaging using JAX~\cite{jax2018github} which consists of a differentiable reimplementation of the SIMUS frequency-domain point-scatterer model~\cite{garcia2022} for the forward RF, a differentiable delay-and-sum (DAS) beamformer, and differentiable image quality objectives (contrast, resolution, coherence and speckle brightness) that can be jointly trained and reported. The element pitch, the transmit and receive apodizations, and the global speed of sound all enter the same computation graph, so the probe geometry along with the software parameters is differentiable as shown in Fig.~\ref{fig:arch}. The main contributions of this work are:
\begin{itemize}
\item \textbf{One differentiable pipeline:} speckle phantoms, a frequency-domain RF simulator, delay-and-sum beamforming and an image-quality objective in one computation graph. 
\item \textbf{Joint optimization with attribution:} the framework optimizes the element pitch, the transmit and receive apodization and the global speed of sound together.
\item \textbf{Transfer to unseen phantoms:} per-element apodization weights overfit the speckle of the tuned phantom, but a smooth cosine parameterization gives a receive gain that transfers with a $+0.037$ gCNR on 14 unseen phantoms.
\item \textbf{A configurable image-quality objective:} a user-weighted sum of differentiable image quality metrics, with an ablation that shows the effect of each metric. 
\end{itemize}

\section{Related Work}
Differentiable frameworks for simulation and beamforming has been getting attention in the recent past. There are differentiable acoustic simulators like j-Wave~\cite{stanziola2023} which is an accelerated k-Wave~\cite{treeby2010kwave} pseudo-spectral solver in JAX and built for full-waveform inversion. Such systems provide gradients of the wavefield with respect to the medium parameters but are not formulated as design tools to optimize a task-level objective. Also, they are computationally heavy and give stiff gradients. On the differentiable beamforming side, there are approaches such as vbeam~\cite{kvalevag2023} and differentiable-beamforming autofocusing~\cite{simson2023,simson2026cmpe} that optimize imaging or medium parameters on already-acquired data, with the probe layout and transmit sequence unchanged. Open toolboxes for differentiable ultrasound are also emerging, e.g.\ \texttt{zea}~\cite{zea} which is a JAX/Keras pipeline for differentiable beamforming and acquisition. The approach in \cite{goudarzi2025oats} learns depth-dependent receive apodization through a differentiable beamformer trained on simulations. Spainhour~et~al.~\cite{spainhour2024} performs transmit encoding (apodization, delays) optimization through a custom DAS beamformer with the Field~II forward model and the design parameter is the transmitted waveform for a fixed probe layout. Dogan~et~al.~\cite{dogan2025coded} optimize the coding matrix of unfocused transmissions with a linear signal model, and van der Meulen~et~al.~\cite{vandermeulen2020coding} design the coding mask of a single-sensor ultrasound imager. Unlike our work, these are general processing frameworks rather than probe-geometry or transmit design tools. For probe-geometry optimization, probe layouts have traditionally been optimized through genetic algorithms and simulated annealing~\cite{austeng2002,trucco1999}, convex relaxations~\cite{diarra2013}, and analytic density-tapered spiral designs for 2-D arrays~\cite{ramalli2015} and more recently with reinforcement learning for sparse-array beamforming~\cite{tamraoui2025drl}. A differentiable design framework for sparse linear probes was proposed in~\cite{urrea2026} that learns a binary mask of selected elements at fixed pitch from a point-spread-function objective. More broadly, learning-based work has targeted the receive processing (learned apodization~\cite{goudarzi2025oats}) and the acquisition scheme through task-adaptive subsampling~\cite{huijben2020dps}. Finally, learning the acquisition through differentiation is established in MRI such as learned sampling masks~\cite{bahadir2020loupe}, pulse sequences~\cite{loktyushin2021}, and hardware-constrained k-space trajectories~\cite{wang2022bjork,wang2023snopy}, and in deep optics~\cite{sitzmann2018,wang2022dO,suarez2025distilling}. A comparison of the proposed approach with respect to differentiable approaches in ultrasound imaging is presented in Table~\ref{tab:related}.

\begin{table}[t]
\caption{Comparison of differentiable ultrasound/imaging work in literature}
\label{tab:related}
\centering\scriptsize
\resizebox{\columnwidth}{!}{%
\begin{tabular}{llll}
\toprule
Work & Design variable & Forward model & HW in loop \\
\midrule
j-Wave~\cite{stanziola2023} & medium (FWI) & wave eq. & no \\
vbeam~\cite{kvalevag2023} & beamformer & none & no \\
Simson~\cite{simson2023} & speed of sound & none & no \\
Spainhour~\cite{spainhour2024} & TX encoding & Field~II & no (fixed) \\
Urrea~\cite{urrea2026} & element on/off & SIMUS-style & discrete \\
OATS~\cite{goudarzi2025oats} & RX apodization & simulation & no \\
Dogan~\cite{dogan2025coded} & TX coding matrix & linear model & no \\
\textbf{Proposed work} & \textbf{All parameters} & \textbf{freq.-domain} & \textbf{continuous} \\
\bottomrule
\end{tabular}}
\end{table}

\section{System Architecture}
\label{sec:arch}
\subsection{Overview}
\label{sec:arch:overview}
\begin{figure*}[t]
\centering
\includegraphics[width=\textwidth]{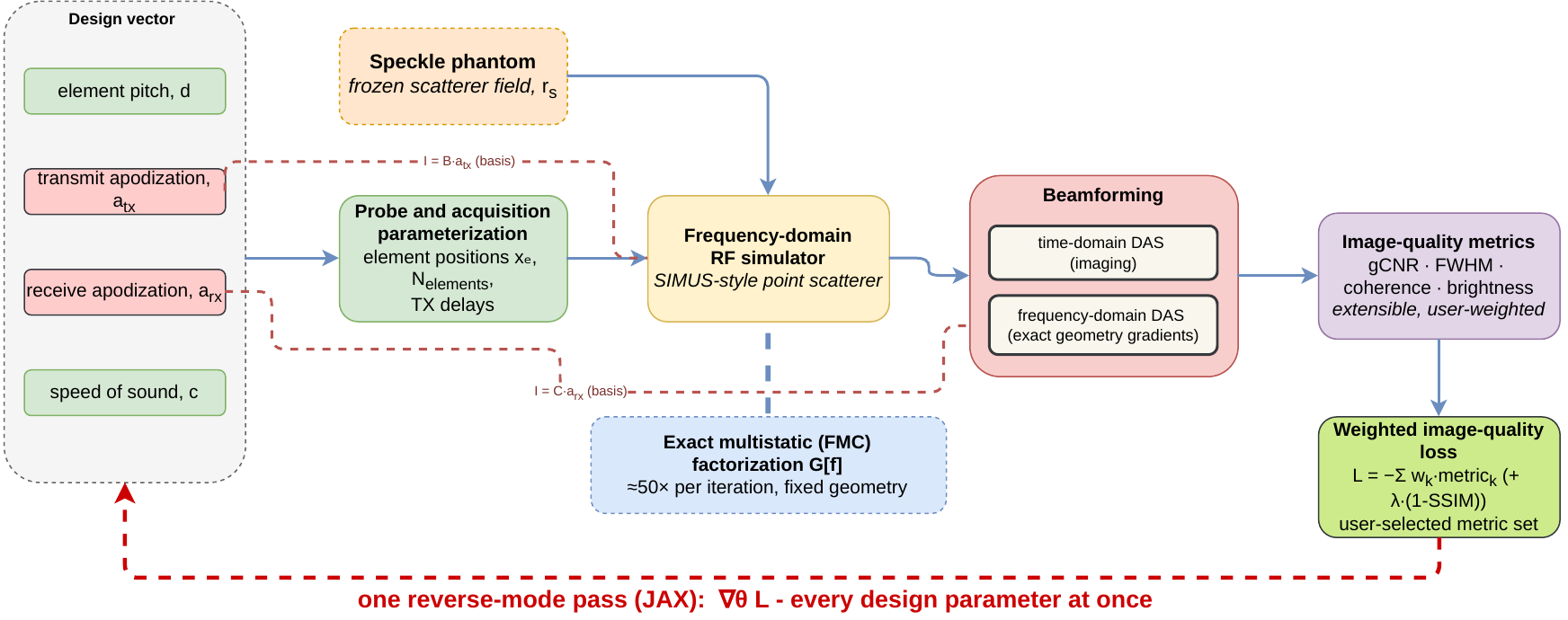}
\caption{The proposed differentiable framework in JAX with the design vector $\theta=\{d,\ \mathbf a_{\mathrm{tx}},\ \mathbf a_{\mathrm{rx}},\ c\}$ which drives a frozen-phantom, frequency-domain RF simulator and a delay-and-sum beamformer feeding user-weighted image-quality objective.}
\label{fig:arch}
\end{figure*}

We formulate plane-wave ultrasound acquisition and probe design as an end-to-end differentiable pipeline. It has to be noted that the framework can be extended to other transmit schemes. The system maps a parameterized hardware/sequence design vector $\boldsymbol{\theta}$ directly to a scalar image-quality objective $\mathcal{Q}$,
\begin{equation}
\mathcal F:\ \theta \longmapsto \mathcal Q, \quad
\theta=\{\,d,\ \mathbf a_{\mathrm{tx}},\ \mathbf a_{\mathrm{rx}},\ c\,\},\quad
\mathcal Q=\sum_{k\in S} w_k\,\tilde Q_k
\label{eq:designvec}
\end{equation}
where $d$ is the element pitch (probe geometry), $\mathbf a_{\mathrm{tx}}$ and $\mathbf a_{\mathrm{rx}}\in\mathbb R^{N}$ are the transmit and receive apodizations, $c$ is the assumed global speed of sound and $\tilde Q_k$ represent the individual image quality metrics (e.g. contrast) with weights $w_k$. Unlike conventional forward-only simulators, every block, from RF synthesis through beamforming to the image-quality score, is written as one differentiable JAX program. A single reverse-mode pass therefore returns the exact gradient $\nabla_\theta \mathcal Q$ with respect to all continuous design parameters simultaneously at a computational complexity which is independent of the parameter dimension. An overview of the proposed approach is shown in Fig.~\ref{fig:arch} and explained in detail in the subsequent sections.
\subsection{Design Parameters}
\label{sec:arch:probe}
The key design parameters are the element pitch, $d$, number of elements, $N$, the transmit apodization, $\mathbf a_{\mathrm{tx}}$, global speed of sound of the medium, $c$ and the receive apodization, $\mathbf a_{\mathrm{rx}}$. It is known that the grating-lobe clutter depends on the fractional pitch $d/\lambda=d \times f_c/c$ ($f_c$ is the transmit frequency) and by sampling the aperture at $d\le\lambda/2$, it can be suppressed. We use the L7-4 probe as a common, reproducible linear array, but the method is not restricted to it: what matters is the fractional pitch $d/\lambda$ set by the center frequency, and a different array only shifts this ratio. Since reducing the pitch at fixed number of elements also shrinks the active aperture $D=Nd$, we divide the pitch parametrization studies into two: a \emph{fixed-$N$} sweep (aperture shrinks) and a \emph{fixed-aperture} sweep in which $N$ grows as $N=\lceil D_0/d\rceil$, where $D_0$ is the fixed reference aperture, so the field of view is preserved. Both the studies are made field of view (FOV) aware so that the image quality is evaluated only inside the valid insonified region $|x|\le D/2$. The \emph{transmit apodization} $\mathbf a_{\mathrm{tx}}$ weights the transmitted signals at each element and the \emph{receive apodization} $\mathbf a_{\mathrm{rx}}$ weights the received signal at each element in the delay-and-sum (DAS) beamformer. The \emph{speed of sound} $c$ decides the round-trip delay and affects the delay correction in the DAS beamformer. For the \emph{transmit/receive apodization}, a weight for each element gives the optimizer $N$ free values, and these weights can over-fit the speckle pattern of one phantom. We thus also use a smooth parameterization, in which the weights are a sum of the $K$ lowest-order cosine terms across the aperture, $\mathbf{a} = g(\boldsymbol{\Phi}\mathbf{w})$, with $\Phi_{nk} = \cos\!\big(\pi k (n + 0.5)/N\big)$ for $k = 0, \dots, K-1$. For transmit, $g$ is the softmax function; for receive, $g(x) = 0.3 + \sigma(x)$, normalized to a mean of 1. With $K = 4$, only aperture-scale shapes are possible, for example a taper, a tilt or an edge roll-off.

\subsection{Background Phantom Image}
The image quality objective is meaningful only if the forward model and beamformer produce fully developed speckle. We generate each phantom which is defined by an echogenicity grayscale image $p\in[0,1]$ (bright~$=$~hyperechoic, dark~$=$~anechoic cyst) together with the lesion regions and point scatterers. The point scatterers are placed on a jittered spatial grid with an average spacing of $0.4\lambda$ for training and $0.2\lambda$ for evaluation to ensure a high scatterer density per resolution cell. The local mean reflectivity is interpolated directly from $p$ and mapped across a contrast parameter $G$ (fixed as $40$~dB) that sets the cyst-to-background reflectivity ratio:
\begin{equation}
r_{\mathrm{mean}}=10^{(G/20)(p-1)}
\end{equation}
placing the cyst $40$~dB below the background, making it effectively anechoic. Each scatterer's amplitude is given a small random jitter around this mean:
\begin{equation}
r=r_{\mathrm{mean}}\,\sqrt{g_1^2+g_2^2},\qquad g_1,g_2\sim\mathcal N(0,\tfrac12)
\end{equation}
where $g_1,g_2$ are independent zero-mean Gaussian draws, so the envelope $r$ is Rayleigh-distributed with unit mean power ($\mathbb{E}[r^2]=r_{\mathrm{mean}}^2$). Packing several scatterers into each resolution cell ($0.2$--$0.4\lambda$ spacing) makes the coherent echo sum circular-complex Gaussian, forming a fully developed speckle. We generate the scatterer field once per phantom and freeze it across all probe designs under comparison. The 15 test phantoms are generated at random, with random cyst sizes and positions. With 3 cysts each, there will be a total of 45 cysts, which is enough to maintain the reported statistics while keeping the Field~II checks and the sweeps manageable. The template $p$ (cropped to the active field of view) serves as our ground-truth reflectivity target $g$ for the whole-image based losses, while the locations of the lesion boundaries and the point scatterer defined in the template provide the signal and background masks for the estimation of the image quality (contrast and resolution).

\subsection{Differentiable RF simulator}
\label{sec:arch:sim}
We model RF signal generation using a frequency-domain point-scatterer formulation similar to SIMUS~\cite{garcia2022}, evaluated over a discrete set of frequencies $f_j$, with wavenumber, $\kappa_j=\frac{2\pi f_j}{c}$, where $c$ denotes the speed of sound. For a scatterer $s$ located at $(x_s,z_s)$ and a linear array element $e$ centered at the lateral position $x_e$, the propagation distance is $r_{se} = \sqrt{(x_s-x_e)^2 + z_s^2}$ and for a uniform linear array, the element position is defined as $x_n=\left(n-\frac{N-1}{2}\right)d$ with pitch $d$. For transmit event $q$, the complex transmit coefficient is defined as
\begin{equation}
A_{q,e}[f_j]= a_{q,e}\exp\left(i\,2\pi f_j \tau_{q,e}\right)
\label{eq:tx_coefficient}
\end{equation}
with transmit apodization weight \(a_{q,e}\) and transmit delay \(\tau_{q,e}\) and since \(\kappa_j c = 2\pi f_j\), this phase is a pure transmit time delay. The one-way response between element $e$ and scatterer $s$ is
\begin{equation}
G_{se}[f_j]= S(r_{se})\exp\left(-\alpha(f_j)r_{se}\right)\exp\left(i\,\kappa_j r_{se}\right)D_{se}[f_j]
\label{eq:propagation_response}
\end{equation}
where $S(r_{se})\!\propto\!1/\sqrt{r_{se}}$ is the 2-D geometric spreading, $\alpha(f_j)$ is the attenuation coefficient, and $D_{se}[f_j]$ is the element directivity. The incident field at scatterer $s$ is the coherent sum over transmit elements
\begin{equation}
P^{\mathrm{inc}}_{q,s}[f_j] = \sum_{e=0}^{N-1} G_{se}[f_j] A_{q,e}[f_j]
\label{eq:incident_field}
\end{equation}
and, assuming weak single scattering and reciprocity, the spectrum at receive element $m$ is
\begin{equation}
Y_{q,m}[f_j]= H[f_j]\sum_s \rho_s \left[\sum_{e=0}^{N-1} G_{se}[f_j]\,A_{q,e}[f_j]\right] G_{sm}[f_j]
\label{eq:received_spectrum_expanded}
\end{equation}
where $\rho_s$ is the reflectivity of scatterer $s$ and $H[f_j]$ denotes the combined pulse-transducer response. The frequency sweep is implemented using the \texttt{lax.scan} in JAX, compiling the forward model into a memory-efficient differentiable kernel, so one reverse-mode pass gives gradients with respect to pitch $d$ (through $x_e$ and $r_{se}$), speed of sound $c$ (through $\kappa_j$), and transmit apodization $a_{q,e}$ (through~\eqref{eq:tx_coefficient}).We compared the proposed simulator against zea toolbox~\cite{zea} on the same L7-4 probe ($N{=}128$), scatterers, and frequency grid (about 500 bins). The zea toolbox builds the full scatterer-by-element-by-frequency response before summing, about 0.19~GB per scatterer, so it ran out of memory beyond a few dozen scatterers on the same machine. By scanning over frequency and chunking the scatterers, the proposed simulator held peak memory near 0.1~GB and simulated a 100{,}000-scatterer phantom in 28~s. For a fixed array geometry, medium parameters, transmit delays, and receive beamforming, the image is linear in $a_q$, so it factorizes as
\begin{equation}
\mathbf{I}_q=\mathbf{B}_q \mathbf{a}_q
\label{eq:tx_basis_factorization}
\end{equation}
where
\begin{equation}
\mathbf{a}_q=\begin{bmatrix} a_{q,0} & a_{q,1} & \cdots & a_{q,N-1} \end{bmatrix}^{\mathrm{T}}
\label{eq:tx_apodization_vector}
\end{equation}
is the transmit apodization vector and the $n$-th column of $\mathbf{B}_q$ is the beamformed contribution of transmit element $n$, including its fixed transmit delay. $\mathbf{B}_q$ is precomputed once for a fixed phantom, array geometry, and transmit-delay configuration, reducing the apodization optimization to a matrix-vector product. When the element pitch $d$ changes, $\mathbf{B}_q$ also changes, so the geometry is differentiated through the simulator itself.
\subsection{Differentiable Beamformer}
\label{sec:arch:beamform}
Two delay-and-sum (DAS) beamformers are used because the optimization variables affect the reconstruction differently. A time-domain DAS beamformer is used on the demodulated IQ data for image formation and gives gradients for receive apodization and speed of sound. A frequency-domain DAS formulation is used only for probe-geometry optimization, where differentiating through fractional-sample interpolation can give unstable phase-dominated gradients. For a pixel at \((x,z)\), let \(d_{\mathrm{TX}}\) denote the transmit propagation distance and let the receive distance from element \(e\) located at \(x_e\) be $d_{\mathrm{RX},e}=\sqrt{(x-x_e)^2+z^2}$. The corresponding total propagation delay is $\tau_e=\frac{d_{\mathrm{TX}}+d_{\mathrm{RX},e}}{c}$, where \(c\) is the assumed speed of sound. The complex beamformed image is then obtained as
\begin{equation}
I(x,z)=\sum_e a_{\mathrm{rx},e}\,\mathrm{interp}\left(\mathrm{iq}_e,\tau_e f_s\right)\exp\left(i\,2\pi f_c\tau_e\right)
\label{eq:das}
\end{equation}
where \(a_{\mathrm{rx},e}\) is the receive apodization weight, \(f_s\) is the sampling frequency, and \(f_c\) is the demodulation carrier frequency. Linear interpolation is used to evaluate the IQ signal at fractional-sample delays, while the exponential term compensates for the carrier phase associated with the propagation delay. When the probe geometry is fixed, the delayed and phase-corrected contribution of each receive channel to every image pixel is also fixed. Consequently, the image depends linearly on the receive-apodization vector and can be written as
\begin{equation}
\mathbf{I}=\mathbf{C}\mathbf{a}_{\mathrm{rx}}
\end{equation}
where the \(e\)-th column of \(\mathbf{C}\) is the complex image contribution from receive element \(e\). The matrix \(\mathbf{C}\) is therefore precomputed once for each fixed geometry, reducing receive-apodization updates to a matrix--vector multiplication. The speed of sound remains differentiable through both the propagation delays \(\tau_e\) and the phase correction in Eq.~\eqref{eq:das}. This precomputed representation cannot be used when the probe pitch is optimized, because changing the pitch changes the element positions and hence the propagation distances. For this geometry gradient, we instead use a frequency-domain DAS beamformer:
\begin{equation}
I(x,z)=\sum_e \sum_j Y_e[f_j]\,\exp\left(-i\,\kappa_j\left(d_{\mathrm{TX}}+d_{\mathrm{RX},e}\right)\right)
\label{eq:fd_das}
\end{equation}
where \(Y_e[f_j]\) is the received spectrum at element \(e\) and \(\kappa_j=2\pi f_j/c\). The frequency-domain formulation avoids the interpolation and takes the geometry gradient directly through the propagation phase.
\subsection{Design Objective}
\label{sec:arch:fc}
The design objective is framed as a differentiable, extensible image-quality score. The objective is a weighted sum over a user-selected set $S$ of figures of merit,
\begin{equation}
\mathcal Q(\theta)=\sum_{k\in S} w_k\,\tilde Q_k(\theta),\qquad w_k\ge 0
\label{eq:multimetric}
\end{equation}
where each $\tilde Q_k$ is any quantity expressible as a differentiable function of the beamformed image and normalized to a common scale so the weights are comparable. In this study the differentiable equivalent of four image quality metrics are implemented: generalized contrast to noise ratio (gCNR)~\cite{rodriguez2020} for contrast, full width half maximum (FWHM) for resolution, coherence factor (CF) for delay alignment (for beamforming) and speckle brightness (SB). Any additional metrics can be added to the objective as long as it is differentiable, and a metric can be excluded by setting its weight to zero. The user selects the subset relevant to the task (e.g.\ contrast alone for lesion detection, contrast $+$ resolution for point-target tasks). Section~\ref{sec:multi} reports the ablation studies for each of these metrics to show its effect on the converged design, guiding the choice of $S$ and the weights. The over-all design loss is
\begin{equation}
L(\theta)=-\mathcal Q(\theta)+\lambda\,\mathcal L_{\mathrm{SSIM}}
\end{equation}
where $L_\mathrm{SSIM}$ denotes the structural similarity loss, $(1-\mathrm{SSIM})$, between the generated image and the conventional DAS beamformed output.

For contrast measurement, the standard $\text{gCNR}$, which calculates the histogram overlap of envelope pixels inside and outside the lesion is employed. Since the discrete histograms have no gradients, the gCNR estimation is made differentiable by replacing the hard histogram bins with Gaussian kernel density estimates (KDE):
\begin{equation}
\tilde{h}_R[b] = \frac{\sum_{i \in R} \exp\left(-\frac{(v_i - b)^2}{2\sigma_h^2}\right)}{\Delta b \sum_{b'} \sum_{i \in R} \exp\left(-\frac{(v_i - b')^2}{2\sigma_h^2}\right)} \qquad R \in \{\mathcal{I}, \mathcal{O}\}
\label{eq:soft_hist}
\end{equation}
\begin{equation}
\text{gCNR}_{\text{soft}} = 1 - \sum_{b} \min\big(\tilde{h}_{\mathcal{I}}[b], \, \tilde{h}_{\mathcal{O}}[b]\big) \Delta b
\label{eq:soft_gcnr}
\end{equation}
Unlike the variance based CNR approaches that distort the dynamic range, the $\text{gCNR}_{\text{soft}}$ directly approximates the true metric~\cite{rodriguez2020}. The differentiable $\text{gCNR}_{\text{soft}}$ and the exact histogram-overlap $\text{gCNR}$ are correlated with a Pearson coefficient of $r = 0.99$. All lesion regions are defined as physical ellipses directly in the phantom coordinates, ensuring no background pixels leak into the lesion masks due to pixel aspect-ratio mismatches. The resolution (we only consider lateral resolution here) is measured on a point target from the lateral intensity profile $p(x)$ in a window around the target as an intensity-weighted main-lobe width,
\begin{equation}
W=\sqrt{\frac{\sum_x w(x)\,(x-\bar x)^2}{\sum_x w(x)}}
\label{eq:mainlobe_width}
\end{equation}
with weights $w(x)=\big(p(x)-\min_x p\big)^2$ and centroid $\bar x=\sum_x w(x)\,x/\sum_x w(x)$. $W$ is differentiable in the pixel values, and the $p^2$ weighting emphasizes the main lobe so it cannot be reduced by merely tapering sidelobes. The conventional $-6$\,dB lateral FWHM, which is non-differentiable, is used only for reporting at evaluation. The coherence factor, CF, is computed from the per-channel, delay and apodization corrected signals $s_e(x)$ (the terms summed by the beamformer)~\cite{mallart1994adaptive,li2003adaptive} is given by
\begin{equation}
\mathrm{CF}(x)=\frac{\big|\sum_{e=1}^{N} s_e(x)\big|^2}{N\sum_{e=1}^{N}\lvert s_e(x)\rvert^2}\in[0,1]
\label{eq:coherence}
\end{equation}
which is a smooth rational function of the complex channel data (differentiable in $c$ through the delays and in $\mathbf a_{\mathrm{rx}}$), averaged over the speckle region. The speckle brightness, SB, used typically as a focus-quality criterion~\cite{nock1989phase}, is the mean log-envelope over the speckle region,
\begin{equation}
SB=\frac{1}{|\Omega_s|}\sum_{x\in\Omega_s}\log\!\big(\lvert I(x)\rvert+\varepsilon\big)
\label{eq:brightness}
\end{equation}
differentiable through the envelope $\lvert I\rvert$ and the logarithmic function. It has to be noted that each image quality metric is evaluated only inside the region appropriate to it, lesion masks for gCNR, a point-target window for FWHM, and the speckle region for coherence and brightness.
\subsection{Image fidelity objective}
\label{sec:arch:fidelity}
Evaluating the image quality metrics (gCNR, Resolution, CF and SB) only on local regions allows the optimizer to boost only the local contrast/resolution metrics by degrading the surrounding speckle texture. To avoid this, an image fidelity check is introduced against the conventional DAS beamformed output within the active field of view. Taking $\hat{I}$ as the normalized, log-compressed reconstruction, we compute the differentiable mean-squared error and structural similarity index over the valid FOV $\Omega$:
\begin{equation}
\begin{aligned}
\mathcal{L}_{\text{MSE}} &= \frac{1}{|\Omega|} \sum_{\Omega} \big(\hat{I}(x, z) - g(x, z)\big)^2 \\
\mathcal{L}_{\text{SSIM}} &= 1 - \text{SSIM}\big(\hat{I},g\big)
\end{aligned}
\label{eq:fidelity_losses}
\end{equation}
These global terms can either replace the image quality metrics entirely or combine with it as $\mathcal{L} = -\mathcal Q(\theta) + \lambda \mathcal{L}_{\text{SSIM}}$ to keep background artifacts in check.

\subsection{Optimization}
\label{sec:arch:res}
Each design problem is optimized using the projected adaptive momentum (Adam)~\cite{kingma2014adam} dynamics over the scalar loss function:
\begin{equation}
\boldsymbol{\theta}_{t+1} = \Pi_{\mathcal{B}} \left( \boldsymbol{\theta}_t - \alpha \, \mathrm{Adam}\big(\nabla_{\boldsymbol{\theta}} \mathcal{L}\big) \right),
\label{eq:projected_adam}
\end{equation}
where $\Pi_{\mathcal{B}}$ denotes projection onto the allowable physical parameter bounds $\mathcal{B}$, and $\nabla_{\boldsymbol{\theta}} \mathcal{L}$ is computed in a single reverse-mode pass. To prevent continuous apodization from introducing non-physical acoustic nulls in the array, element weights follow a bounded parameterization, $a_e \propto a_{\min} + \sigma(w_e)$, ensuring the active aperture remains filled. For the discrete transmit variant, we apply a hard threshold to the converged weights in a post-hoc manner. We also independently verify low-dimensional parameters (such as pitch) via exhaustive grid sweeps. Finally, all converged configurations and loss ablations are reported on held-out test phantoms using the exact, non-differentiable image-quality metrics. A detailed overview of the proposed algorithmic framework is shown in Algorithm~\ref{alg:codesign}.

\begin{algorithm}[t]
\caption{Gradient-based co-design of probe geometry and acquisition}
\label{alg:codesign}
\begin{algorithmic}[1]
\Require phantom set $\mathcal{P}$, bounds $\mathcal{B}$, metric weights $\{w_k\}_{k\in S}$, SSIM weight $\lambda$, step $\alpha$, iterations $T$
\Ensure optimized design $\boldsymbol{\theta}^\star=\{d,\mathbf a_{\mathrm{tx}},\mathbf a_{\mathrm{rx}},c\}$
\State Initialize $\boldsymbol{\theta}$ and the Adam state
\For{each phantom $p\in\mathcal{P}$} \Comment{fixed geometry}
    \State precompute TX/RX basis matrices $\mathbf B_p,\mathbf C_p$
\EndFor
\For{$t=1$ to $T$}
    \State $\mathcal{L}_{\mathrm{cum}}\gets 0$
    \For{each phantom $p\in\mathcal{P}$}
        \State $\mathbf Y \gets \textsc{Simulate}(\boldsymbol\theta; p)$ \Comment{freq.-domain RF}
        \State $\mathbf I \gets \textsc{Beamform}(\mathbf Y;\boldsymbol\theta)$ \Comment{$\mathbf B_p\mathbf a_{\mathrm{tx}}$, $\mathbf C_p\mathbf a_{\mathrm{rx}}$, DAS}
        \State $\mathcal{Q}\gets \sum_{k\in S} w_k\,\tilde Q_k(\mathbf I)$ \Comment{gCNR, FWHM, coherence, brightness}
        \State $\mathcal{L}_{\mathrm{cum}} \gets \mathcal{L}_{\mathrm{cum}} - \mathcal{Q} + \lambda\,\mathcal{L}_{\mathrm{SSIM}}(\mathbf I)$
    \EndFor
    \State $\mathbf g \gets \nabla_{\boldsymbol\theta}\mathcal{L}_{\mathrm{cum}}$ \Comment{one reverse-mode pass}
    \State $\boldsymbol\theta \gets \Pi_{\mathcal{B}}\big(\boldsymbol\theta - \alpha\,\mathrm{Adam}(\mathbf g)\big)$
    \If{$d$ (geometry) changed}
        \State recompute $\mathbf B_p,\mathbf C_p$ for all $p$
    \EndIf
\EndFor
\State \Return $\boldsymbol\theta^\star$
\end{algorithmic}
\end{algorithm}

\subsection{Implementation Details}
The L7-4 transducer array with probe geometry, $N = 128$ elements, $298~\mu$m pitch, element width $w = 0.9\,d$ (so the kerf scales with original pitch of $268~\mu$m), and $67\%$ fractional bandwidth, with transmit center frequency $f_c = 4$~MHz, sampled at $f_s = 4f_c = 16$~MHz is used as the reference probe for this work. The frequency-domain forward simulator evaluates the frequency sub-band of $[0, 2f_c]$ where the combined pulse--probe spectral weighting $w_f = |P_{\mathrm{pulse}}(\omega_f)|\,|P_{\mathrm{probe}}(\omega_f)|^2$ exceeds $-60$~dB under a Gaussian probe response model. For gradient optimization, we make each image metric a smooth function of the beamformed image as explained in Sec~\ref{sec:arch:fc}. Element apodizations are constrained as $a_e = a_{\min} + \sigma(w_e)$ with $a_{\min} = 0.30$ ($a_e \in [0.30, 1.30]$) to keep all channels active and avoid acoustic holes in the aperture. Transmit weights are normalized to unit sum, receive weights to unit mean and the discrete transmit aperture is obtained by thresholding converged continuous weights. For single-parameter sweeps, we use Adam with a learning rate of $0.01$. For joint co-design, we use Adam with an exponentially decaying learning rate starting at $0.03$. Both optimizations run for $150$--$250$ iterations, with projection onto the physical bounds $\mathcal{B}$. The composite SSIM loss uses a regularization weight of $\lambda = 0.15$. Since the channel-data basis are precomputed once per medium across $N$ forward simulations, the subsequent reverse-mode gradient steps reduce to dense matrix-vector products executing in milliseconds, requiring $\approx 0.27$~GB of video RAM (VRAM) per phantom at $N = 198$. To preserve the native aperture of 38.1~mm, it is sampled at optimal pitch $\lambda/2$ ($d = 192.5~\mu\text{m}$) and $N$ is calculated as $N = \lceil D_0/d \rceil = 198$ elements.

\section{Experiments}
\label{sec:experiments}

\subsection{Simulator validation}
\label{sec:fieldii}
\begin{figure}[t]
\centering
\includegraphics[width=\linewidth]{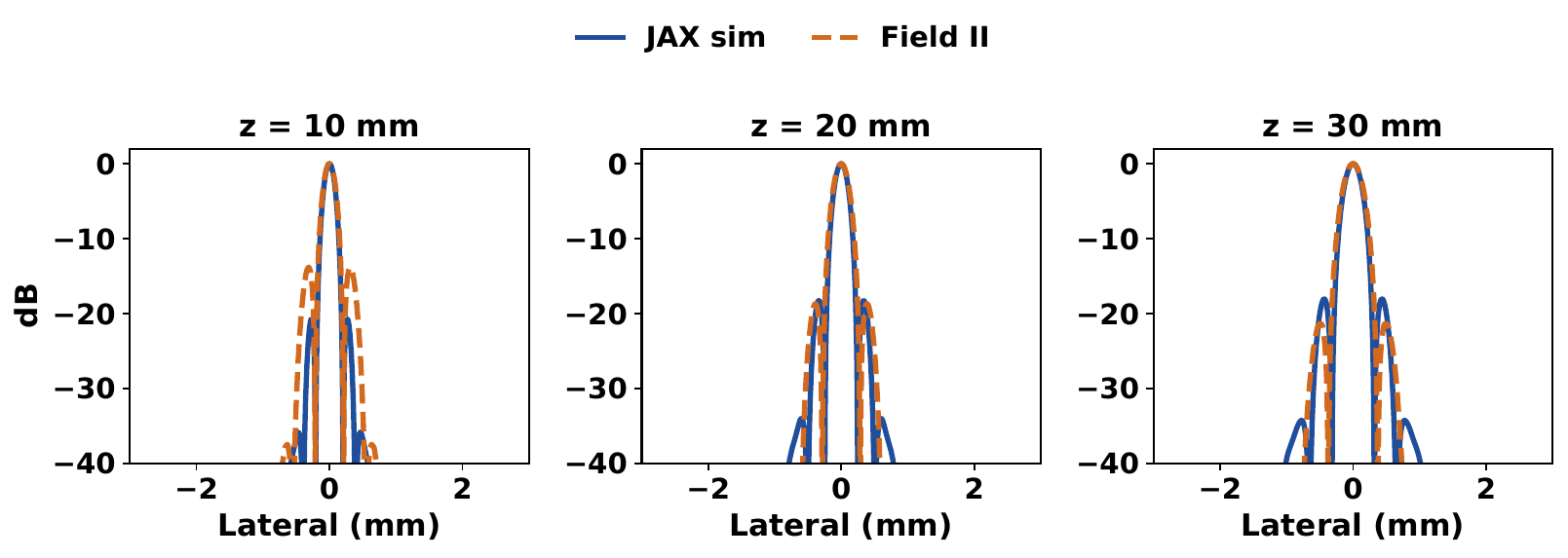}
\caption{Lateral point-spread-function comparison against Field~II on a matched L7-4 geometry and scatterer field: the FWHM agrees within $\sim$10\% across depth.}
\label{fig:fieldii}
\end{figure}

The proposed forward model is verified against the  Field~II~\cite{jensen} whose spatial impulse response model is independent of our frequency-domain formulation. The RF data from both the simulators are beamformed with the same delay-and-sum pipeline and the point-spread functions are compared on a matched L7-4 geometry, pulse, and common scatterer field. The lateral FWHM matches to within $\sim$10\% across depth (ours $0.228/0.282/0.363$~mm vs Field~II $0.250/0.320/0.405$~mm at $10/20/30$~mm), with the main lobes co-registered at $x=0$ as shown in Fig.~\ref{fig:fieldii}. The point scatterer from the proposed simulator is narrower than that of Field~II at every depth, since we leave out the full element response and the elevation aperture to keep the simulator fast and differentiable. The ablation in Table~\ref{tab:sim_ablation} confirms that the directivity and obliquity terms pull our point scatterers back towards Field~II, and what is left over is the broadening we dropped, which widens with the depth. This offset does not limit the design results, because every design is compared on the same simulator and hence a constant offset cancels out.
\begin{table}[t]
\centering
\caption{RF-simulator ablation: lateral FWHM (mm) at 10/20/30~mm as each physics term is toggled. The terms we model move the point toward Field~II; what remains is the elevation and element-response broadening we do not model.}
\label{tab:sim_ablation}
\setlength{\tabcolsep}{5pt}
\begin{tabular}{lccc}
\toprule
Configuration & 10~mm & 20~mm & 30~mm \\
\midrule
Field~II (reference)         & 0.250 & 0.320 & 0.405 \\
Full model (ours)            & 0.235 & 0.308 & 0.354 \\
\quad directivity off        & 0.195 & 0.279 & 0.337 \\
\quad obliquity off          & 0.213 & 0.294 & 0.346 \\
\quad 3-D spreading ($1/r$)  & 0.247 & 0.315 & 0.359 \\
\bottomrule
\end{tabular}
\end{table}

\subsection{Single-parameter studies}
\label{sec:single}
\begin{figure*}[t]\centering
\includegraphics[width=0.9\textwidth]{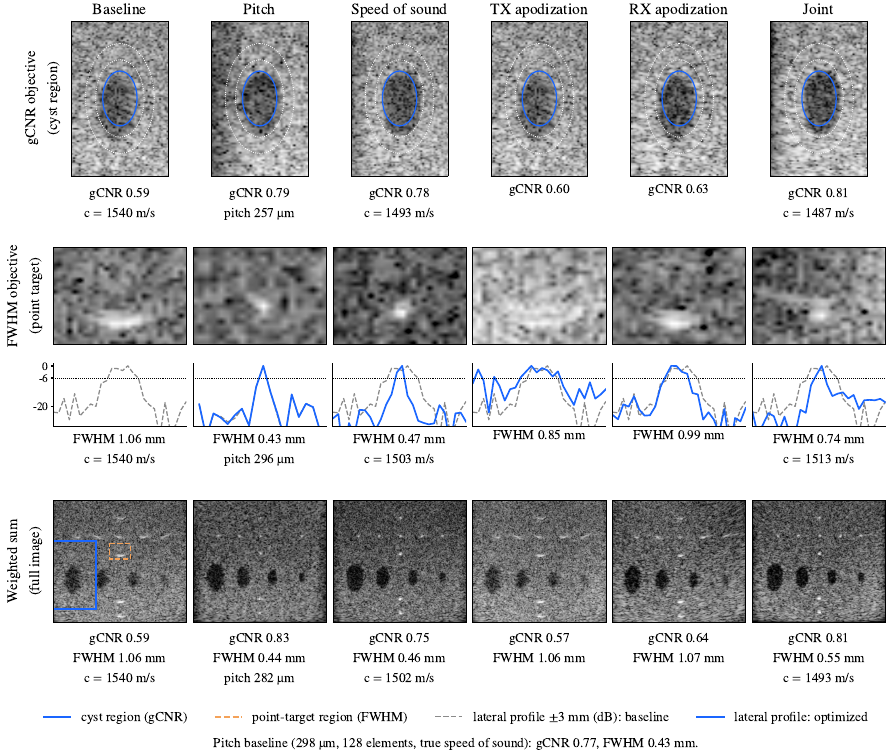}
\caption{Effect of each system parameter on the QA phantom. Each column optimizes one parameter (pitch, speed of sound, TX and RX apodization and the combined version) and each row uses one objective: gCNR (shown on the cyst region), FWHM (shown on the point-target region, with the lateral profile through the point target) and a weighted sum of gCNR, FWHM, coherence and brightness (full image). Each panel shows the best of five runs from different start points and each zoomed view is normalized to its own maximum. All images use a 40~dB dynamic range. The values below each panel are exact metrics of the full image (which is the mean of all cysts and all point targets).}
\label{fig:study_bmodes}
\end{figure*}

In this study, we optimize one design parameter at a time for each image-quality objective, and all other parameters remain constant. The transmit is one plane wave at normal incidence and the apodization and speed-of-sound (SOS) studies use the optimized geometry: the native 38.1~mm aperture with a 192.5~$\mu$m pitch, which gives 198 elements. Fig.~\ref{fig:study_bmodes} shows the results on a quality-assurance (QA) phantom with four anechoic cysts of decreasing size and nine point targets. Each column of Fig.~\ref{fig:study_bmodes} shows one parameter, and each row shows the image region that its objective measures: a cyst and its background ring for the gCNR, a point target and its lateral profile for the FWHM, and the full image for the weighted sum. Each panel is the best of five runs from different start points, and Table~\ref{tab:lever_metric} gives the gCNR and FWHM of each panel. 
\begin{table}[t]
\centering
\caption{Image quality on the QA phantom, as cyst gCNR\,/\,lateral FWHM (mm). Rows: optimized parameter. Columns: objective of the optimization. Each value is the best of five runs, selected by the objective of its column.}
\label{tab:lever_metric}
\setlength{\tabcolsep}{3pt}
\resizebox{\columnwidth}{!}{%
\begin{tabular}{lccccc}
\toprule
Parameter & gCNR & FWHM & CF & SB & Weighted sum \\
\midrule
\textit{Baseline} & \multicolumn{5}{c}{\textit{0.59\,/\,1.06 \ (1540~m/s, rect.\ apod., opt.\ geometry)}} \\
\midrule
Pitch          & 0.79/0.48 & 0.78/0.43 & 0.67/0.66 & 0.75/0.42 & 0.83/0.44 \\
Speed of sound & 0.78/0.50 & 0.75/0.47 & 0.53/1.30 & 0.51/1.26 & 0.75/0.46 \\
TX apod.       & 0.60/1.06 & 0.51/0.85 & 0.59/1.06 & 0.56/1.06 & 0.57/1.06 \\
RX apod.       & 0.63/1.13 & 0.60/0.99 & 0.59/1.06 & 0.65/1.11 & 0.64/1.07 \\
Joint          & 0.81/0.59 & 0.50/0.74 & 0.47/1.09 & 0.59/1.04 & 0.81/0.55 \\
\bottomrule
\end{tabular}}
\end{table}

\textbf{Probe geometry:} The element pitch is the differentiable parameter in this study and the gradient of pitch comes from the automatic differentiation through the frequency-domain simulator. The gradient includes the effect of the element position and of the element width. With a constant number of elements ($N=128$), a smaller pitch makes the aperture smaller, which reduces the grating lobes but worsens the lateral resolution, and the optimizer must balance these two effects. In Fig.~\ref{fig:study_bmodes} (pitch column), the five runs start at pitch values from 180~$\mu$m to 298~$\mu$m and stop at different stationary points. The best result for gCNR is at 257~$\mu$m ($d/\lambda = 0.69$, gCNR $0.77 \to 0.79$), and the best result for FWHM stays near the initial pitch (296~$\mu$m, FWHM $0.43 \to 0.43$~mm). At a constant element count, the pitch alone thus gives only a small improvement. To separate the pitch from the aperture size, we compare two configurations as $d \to \lambda/2$: in the first, the number of elements stays at $N=128$, so the aperture becomes smaller and in the second, the aperture stays constant, so the number of elements increases as $N = \lceil D_0/d \rceil$. Both configurations increase the contrast (gCNR $0.55 \to 0.66$). With $N=128$, the lateral resolution becomes worse (FWHM $0.33 \to 0.48$~mm), but with a constant aperture and 198 elements it does not change (FWHM $0.33 \to 0.31$~mm). Over 15 phantoms, the optimized geometry increases the mean cyst gCNR by $+0.111$ ($0.552 \to 0.663$). A second sweep on 15 phantoms, with a constant aperture of 38.1~mm at 4~MHz, confirms this result: the mean gCNR increases from $0.60$ at $d/\lambda = 1.0$ to $0.79$ at $d/\lambda = 0.5$, and the FWHM stays at about 0.56~mm for all pitch values because the aperture does not change. The proposed simulator thus gives the trade-off between image quality and the pitch before the manufacture of a probe.\\
\textbf{Transmit apodization:} Because the wave propagation is linear in the transmit excitation, the beamformed image is $\mathbf{I} = \mathbf{B}\mathbf{a}_{\mathrm{tx}}$, and a precomputed basis $\mathbf{B}$ gives exact gradients. We compare two parameterizations of the weights: one free weight for each element, $\mathbf{a}_{\mathrm{tx}} = \mathrm{softmax}(\mathbf{w})$, and a smooth profile of four cosine terms. With one free weight for each element, the learned weights are irregular, and only about 94 of the 198 elements stay effectively active. These weights increase the gCNR on the tuned phantom ($+0.103$) but decrease it on the 14 unseen phantoms ($-0.082$). With smooth weights, the gain is small on the tuned phantom ($+0.009$) and negative on unseen phantoms ($-0.034$), so neither parameterization transfers to unseen cases. A paired test on 27 held-out lesions gives the same result: relative to uniform excitation, the learned continuous weights change the gCNR by $+0.017$ (Wilcoxon signed-rank $p = 0.66$), and a thinned aperture of 41 of the 128 elements of the native probe, made from the learned weights with a fixed threshold, changes it by $-0.006$ ($p = 0.58$). The transmit gain thus comes from the speckle and the cyst positions of the tuned phantom, not from a better aperture. It has to be noted that, the effect of transmit apodization might be more relevant to multi-angle plane waves which is beyond the scope of this work.\\
\textbf{Receive apodization:} The receive weights also enter delay-and-sum beamforming linearly, as $\mathbf{I} = \mathbf{C}\mathbf{a}_{\mathrm{rx}}$. We keep the weights positive and bounded, so all channels stay active. With one free weight for each element, the weights fit the speckle of the tuned phantom: they increase the gCNR on it ($+0.091$) but decrease it on unseen phantoms ($-0.026$). With smooth weights (four cosine terms), the gain is $+0.064$ on the tuned phantom and $+0.037$ on unseen phantoms. A smooth receive taper thus gives a small contrast gain that transfers to unseen phantoms.\\
\textbf{Speed of sound:} An incorrect speed of sound defocuses the beamformer on all channels, so a correction must give a gain that transfers to other images of the same medium. We simulate the RF data at a true speed of 1500~m/s, start the beamformer at 1540~m/s, and let the optimizer change the scalar parameter $c$. Of the 45 runs on 15 phantoms, from initial values of 1460, 1540 and 1580~m/s, 41 stop within 20~m/s of the true value, with a median error of 6~m/s. On the QA phantom (Fig.~\ref{fig:study_bmodes}), the speed of sound alone gives the largest change of all parameters: gCNR $0.59 \to 0.78$ and FWHM $1.06 \to 0.47$~mm. Over 45 test lesions, the correction gives a paired gCNR improvement of $\Delta = +0.053$ (95\% CI $[+0.029, +0.078]$, Wilcoxon $p = 1.2 \times 10^{-4}$), and 73\% of the lesions show a measurable gain. This study assumes a homogeneous speed of sound and a study of heterogeneous media will be attempted as future work.

\subsection{Multi-parameter co-design}
\label{sec:multi}
With the probe geometry at its optimum, the framework optimizes the transmit apodization, the receive apodization and the speed of sound together. The apodizations enter through their exact basis factorizations ($\mathbf{I}=\mathbf{B}\mathbf{a}_{\mathrm{tx}}$, $\mathbf{I}=\mathbf{C}\mathbf{a}_{\mathrm{rx}}$) and the speed of sound through the propagation delays, so one backward pass gives the gradients of all parameters with no separate search. Because each parameter is differentiable, we can also find the physical source of each improvement. On unseen cysts, the speed-of-sound correction gives the largest gain ($\Delta\mathrm{gCNR} = +0.146$), the receive apodization adds $+0.028$, and the coupling between the parameters adds $+0.001$. Transmit apodization gives no significant gain, so we keep it uniform, and the total gain of $+0.175$ comes from the speed of sound and the receive weights, $\{c, \mathbf{a}_{\mathrm{rx}}\}$. On the QA phantom (Fig.~\ref{fig:study_bmodes}, joint column), the joint optimization gives the highest gCNR (0.81), but for the FWHM objective its result (0.74~mm) is worse than that of the speed of sound alone (0.47~mm). Since the gCNR uses only the cyst masks and the FWHM only the point targets, unconstrained element weights can improve these local values with a sparse aperture while the image becomes worse in other regions. Bounding the receive weights keeps the aperture filled and prevent most of this behavior. A light SSIM anchor (weight 0.15) does not stop it, and a weight of 3 or more stops it but then almost freezes the weights.\\
\textbf{Metric ablation (image-quality objective):} The objective in Eq.~\eqref{eq:multimetric} accepts any weighted sum of differentiable metrics, so the framework does not use one fixed definition of image quality. Each metric moves the optimization toward the property that it measures: the gCNR increases the separation between a lesion and its background, the FWHM decreases the main-lobe width of a point target, the speckle brightness increases the backscatter level, and the coherence factor keeps a filled, phase-aligned aperture and decreases the grating-lobe stripes that the region-based metrics can cause. No single metric is optimal for all tasks, and different parameters respond to different metrics as shown in Table~\ref{tab:lever_metric}. For example, the FWHM objective decreases the gCNR with transmit apodization ($0.51$) and in the joint optimization ($0.50$). On the QA phantom, the CF and SB objectives do not improve the image, so Fig.~\ref{fig:study_bmodes} does not show them and Table~\ref{tab:lever_metric} gives their values. With the CF objective, all five speed-of-sound runs stop at 1556 to 1558~m/s from all start points. With one plane wave, the CF thus has a constant bias of about $+57$~m/s from the true value (1500~m/s), and the image becomes worse (gCNR $0.53$, FWHM $1.30$~mm). This bias comes from the single plane wave, not from the CF itself. It is shown in sec.~\ref{sec:cubdl}, that the multi-angle plane-wave compounding removes the bias. With the SB objective, the runs stop at different values from 1539 to 1575~m/s and hence the SB gives no stable speed of sound.\\
\textbf{Combined objective:} The weighted-sum row of Fig.~\ref{fig:study_bmodes} optimizes four metrics together as one objective, $\mathcal{Q}=w_{\mathrm{c}}\,\mathrm{gCNR}+w_{\mathrm{r}}(-W)+w_{\mathrm{coh}}\,\mathrm{CF}+w_{\mathrm{b}}\,B$, with a light SSIM anchor. The metrics have different numerical scales (gCNR about 0.8, main-lobe width $W$ about 0.5~mm, coherence factor about $10^{-2}$, brightness a few units), so we select the weights such that no metric has a larger effect only because of its scale. This work uses $(w_{\mathrm{c}}, w_{\mathrm{r}}, w_{\mathrm{coh}}, w_{\mathrm{b}}) = (1, 0.3, 30, 0.3)$ and an SSIM weight of 0.15. These weights are a practical starting point that a user can change per task; for example, a higher $w_{\mathrm{coh}}$ decreases clutter. On the QA phantom, the weighted sum gives a high gCNR and a narrow main lobe together in the joint optimization (gCNR 0.81, FWHM 0.55~mm, Table~\ref{tab:lever_metric}).
\begin{figure}[t]
\centering
\includegraphics[width=0.9\columnwidth]{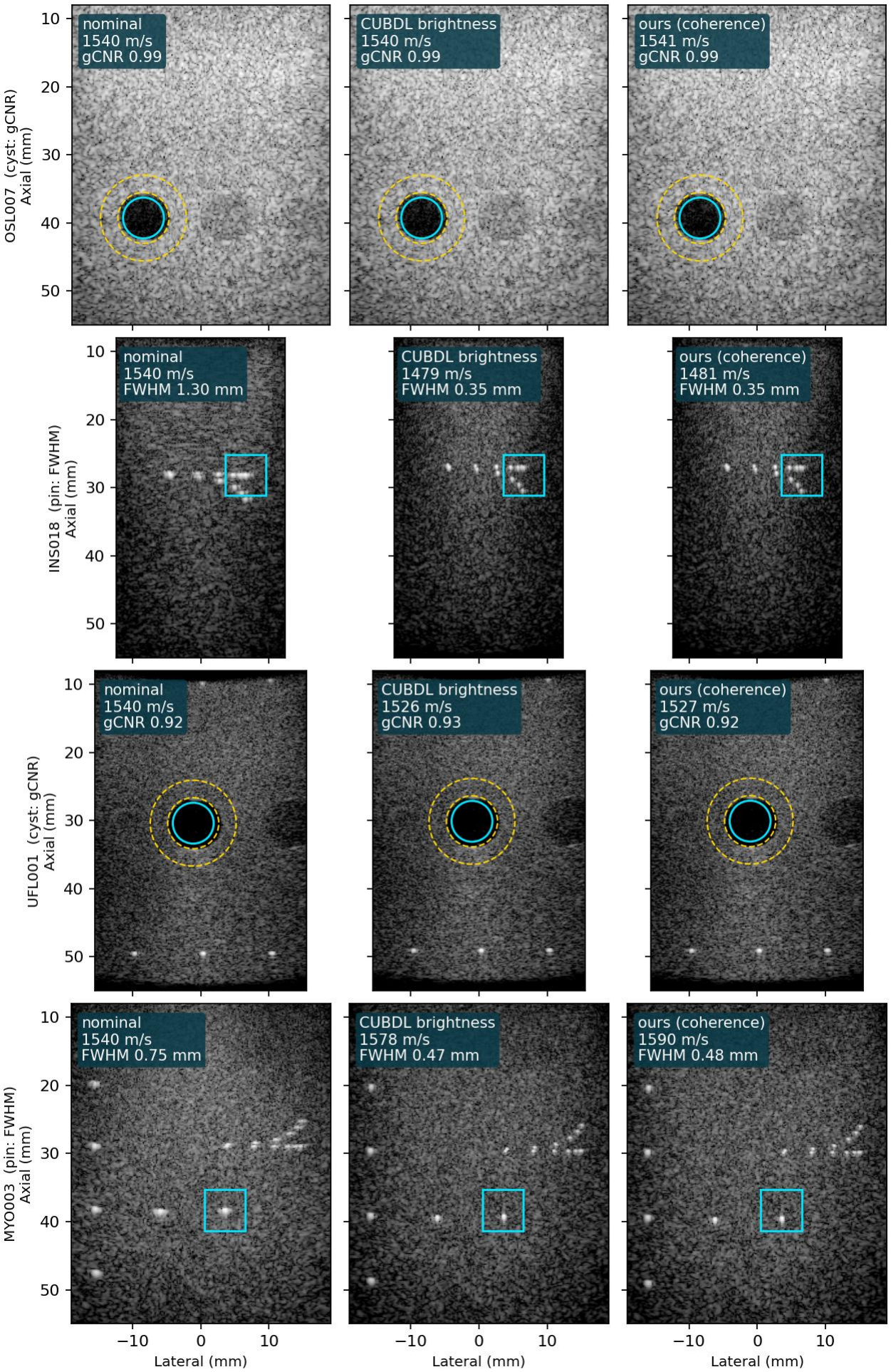}
\caption{Beamformed B-mode images ($-50$~dB) of four CUBDL acquisitions (rows). The columns use the nominal speed of sound (1540~m/s), the speed of sound from CUBDL, and the speed of sound from our optimizer, all with the same operator. Cyan box: point target (FWHM). Cyan and yellow circles: cyst inside and background (gCNR). The corrected speed of sound makes the point targets and the cyst boundaries sharper.}
\label{fig:cubdl_bmode}
\end{figure}
\subsection{Experimental validation: speed of sound estimation on CUBDL}
\label{sec:cubdl}
In the in-silico study of Sec.~\ref{sec:single} we recovered a known speed of sound under a matched model. Here we test the same differentiable beamformer on real recorded channel data, using the Challenge on Ultrasound Beamforming with Deep Learning (CUBDL)\cite{hyun2021cubdl,bell2020cubdl,cubdl_dataport2019} as an independent, peer-reviewed benchmark. For each of the $49$ plane-wave phantom acquisitions from the five CUBDL institutions (OSL, UFL, MYO, EUT, INS), the channel data were loaded through CUBDL loaders, which provide each institution's probe geometry, transmit angles and per-transmit start times. The data were beamformed with our delay and sum beamformer using CUBDL's delay conventions. Each steered transmit is individually delay-and-sum beamformed and the resulting images are coherently compounded across all the transmit angles, the global speed of sound is then optimized through this compounded reconstruction. The global speed of sound is the free parameter optimized by the differentiable beamformer over a homogeneous speckle region. Using CUBDL's speckle brightness objective, our recovered speed of sounds closely matched published benchmarks with mean absolute error of $19.4$\,m/s, $r=0.80$ and near exact on well conditioned linear arrays (OSL: $3.5$\,m/s; UFL: $4.7$\,m/s). Because speed of sound errors primarily cause receive aperture phase errors, we instead optimized the CF over speckle as in Eq.~\eqref{eq:coherence}. This improved mean error to $16.6$\,m/s (median $8.1$\,m/s, $r=0.82$). The coherence-based correction sharpened lateral resolution (mean FWHM: $0.83 \to 0.64$\,mm across $40/49$ acquisitions) with negligible effect on contrast $\Delta\mathrm{gCNR} = +0.01$. The results are presented in Fig.~\ref{fig:cubdl_bmode}.

\section{Discussion}
\label{sec:discussion}
The proposed framework makes image quality a differentiable function of the probe, acquisition and receive parameters. One reverse-mode pass optimizes the full design vector, puts the contrast gain into the physically meaningful parameters (probe geometry and speed of sound), and ranks the other parameters by their gradients. The optimizer recovers the known grating-lobe bound $d \le \lambda / (1 + \sin\theta_{\max})$ which  validates the framework. Two lessons generalize beyond ultrasound: the forward model must produce not only physically accurate images, but also reliable gradients, as an imaging-adequate beamformer may still mislead optimization of parameters. Further, jointly optimizing many heterogeneous parameters can be counterproductive: high-dimensional parameters with little physical influence may overfit to the training media. We therefore validate each parameter alone and then use the results as constraints in the joint optimization. Since the apodization and the global speed of sound are software parameters, a deployed system could re-tune them on the fly to whatever objective the operator cares about, using the same framework.\\
\textbf{Coherence factor and transmit focusing:} With one plane wave, the CF objective does not find the true speed of sound. On the six phantoms, its maximum is at 1567~m/s on average (1545 to 1600~m/s, true value 1500~m/s) for all F-numbers from 1 to 2, and the bias is largest near the probe. It has to be noted that one unfocused plane wave insonifies the full aperture width $D$, so the speckle coherence length is only about $\lambda z / D$ across the receive aperture (0.13~mm at 13~mm depth, below the 0.19~mm pitch), and the CF is almost flat (mean $0.009$, near $1/N$)~\cite{montaldo2009}. Coherently compounding steered plane waves gives a synthetic transmit focus, and the linear per-element basis synthesizes these transmits exactly, with no new simulation. With 5 angles ($\pm 10^\circ$) or 11 angles ($\pm 15^\circ$), the CF maximum moves to the true speed of sound on all six phantoms and the mean CF increases to $0.032$ and $0.062$, while the gCNR maximum stays at the true speed for all angle sets. This result matches with the CUBDL experiment (Fig.~\ref{fig:cubdl_bmode}), which uses compounded plane waves. The correct objective thus depends on the acquisition: a coherence-based objective needs transmit focusing, but a contrast-based objective such as gCNR does not.\\
\textbf{Limitations:} The probe-geometry optimization is a simulation study, because a built array has a fixed pitch and fabricating a separate array for every pitch is impractical. It is instead meant to guide the probe designer to find a suitable pitch before fabrication, so the accuracy of the simulator is important, and we validate the forward model against Field~II (Sec.~\ref{sec:fieldii}). The forward model assumes a homogeneous medium, a 1-D linear array, and a fixed kerf ratio (element width $0.9\,d$) as the pitch varies, and it excludes frequency-dependent attenuation and channel noise, therefore adding these factors to the framework is a direction for future studies focusing on penetration depth and signal-to-noise ratio. We also transmit a single unsteered plane wave, which makes the effect of transmit apodization small, because a flat wavefront insonifies the medium uniformly and confines the transmit weights to the aperture edges. Using steered and compounded plane waves would break this symmetry, creating directional clutter that a transmit apodization could reduce, and since the proposed framework already addresses this, extending the optimization to this is a natural next step.

\section{Conclusion}
\label{sec:conclusion}
We present an end-to-end differentiable pipeline for plane-wave ultrasound that makes the co-design of transducer hardware and acquisition a gradient-based optimization to achieve optimized image quality. One backward pass optimizes four parameters, $\boldsymbol{\theta} = \{d, \mathbf{a}_{\mathrm{tx}}, \mathbf{a}_{\mathrm{rx}}, c\}$, together and ranks their practical value. A pitch near $\lambda/2$ and the speed-of-sound correction give contrast gains that transfer to unseen phantoms ($+0.111$ and $+0.053$ cyst gCNR), but transmit apodization gives no gain that transfers and receive apodization gives only a small gain on the optimized array. The simulator agrees with the lateral PSFs of Field~II within about 10\%, and the same differentiable operator finds the global speed of sound of the 49 real CUBDL acquisitions in agreement with the published corrections and the phantom speed of sound. Because the per-element basis gives steered plane waves exactly, the framework can also design transmit sequences, such as the steering angles, with no new simulation. A differentiable relation between image quality and acoustic design thus gives a principled and efficient method for hardware and sequence co-design.

\section*{Reproducibility}
Code is available at \url{https://github.com/sdehury34/jax-ultrasound-codesign} and a publicly available deployment of the framework is available at \url{https://pulseecho.in/demos/diffUS/}.

\section*{Acknowledgment}
The authors thank Dr. Zainul Charbiwala and Tricog Health for the technical discussions. This work was supported by the Ministry of Education, Singapore, under the Academic Research Fund Tier 1 (GMS 1052), the SIT-SINGA Research Scholarship and Tricog Health. Generative AI tools were used for brainstorming and language editing. All technical claims, experiments, and code were authored and verified by the authors.

\bibliographystyle{IEEEtran}
\bibliography{refs}

\end{document}